\documentclass[prd,twocolumn,superscriptaddress,floatfix,nopacs,preprintnumbers,nofootinbib]{revtex4}
\usepackage[utf8]{inputenc}

\usepackage[caption = false]{subfig}
\usepackage[normalem]{ulem}
\usepackage{xcolor}

\usepackage{mathrsfs}
\usepackage{amssymb,bm}
\usepackage{amsmath}
\usepackage{mathtools}
\usepackage{slashed}
\usepackage{graphics}
\usepackage{graphicx}

\newcommand{\gev}{\mathrm{GeV}}
\newcommand{\fm}{\mathrm{fm}}

\usepackage[breaklinks,colorlinks,citecolor=citcolor,urlcolor=blue,linkcolor=lcolor]{hyperref}

\definecolor{lcolor}{rgb}{0.5,0,0}
\definecolor{citcolor}{rgb}{0,0.3,0.0}

\newcommand{\as}{\alpha_\mathrm{s}}
\newcommand{\der}{\mathrm{d}}
\newcommand{\be}{\begin{equation}}
\newcommand{\ee}{\end{equation}}
\newcommand{\bea}{\begin{eqnarray}}
\newcommand{\eea}{\end{eqnarray}}

\begin{document}

\author{Adrian Dumitru}
\email{adrian.dumitru@baruch.cuny.edu}
\affiliation{Department of Natural Sciences, Baruch College, CUNY, 17 Lexington Avenue, New York, NY 10010, USA}
\affiliation{The Graduate School and University Center, The City University of New York, 365 Fifth Avenue, New York, NY 10016, USA}
\author{Heikki Mäntysaari}
\email{heikki.mantysaari@jyu.fi}
\affiliation{
Department of Physics, University of Jyväskylä,  P.O. Box 35, 40014 University of Jyväskylä, Finland
}
\affiliation{
Helsinki Institute of Physics, P.O. Box 64, 00014 University of Helsinki, Finland
}
\author{Risto Paatelainen}
\email{risto.sakari.paatelainen@cern.ch}
\affiliation{
Helsinki Institute of Physics, P.O. Box 64, 00014 University of Helsinki, Finland
}
\affiliation{
Department of Physics, University of Helsinki, P.O. Box 64, 00014 University of Helsinki, Finland
}

\title{Color charge correlations in the proton at NLO: beyond geometry based intuition}

\preprint{HIP-2021-13/TH}
\begin{abstract}
Color charge correlators provide fundamental information about the proton structure. In this Letter, we evaluate numerically two-point color charge correlations in a proton on the light cone including the next-to-leading order corrections due to emission or exchange of a perturbative gluon.
The non-perturbative valence quark structure of the proton is modelled in a way consistent with high-$x$ proton structure data.
Our results show that the correlator exhibits startlingly non-trivial behavior at large momentum transfer or central impact parameters, and that the color charge correlation depends not only on the impact parameter but also on the relative transverse momentum of the two gluon probes and their relative angle. Furthermore, from the two-point color charge correlator, we compute the dipole scattering amplitude. Its azimuthal dependence differs significantly from a impact parameter dependent McLerran-Venugopalan model based on geometry.
Our results also provide initial conditions for Balitsky-Kovchegov evolution of the dipole scattering amplitude. These initial conditions depend not only on the impact parameter and dipole size vectors, but also on their relative angle and on the light-cone momentum fraction $x$ in the target. 
\end{abstract}

\maketitle

\section{Introduction}

Revealing the proton and nuclear structure at high energies, or equivalently at small momentum fraction $x$, is one of the major tasks of the planned future nuclear deep inelastic scattering (DIS) facilities such as the Electron-Ion Collider (EIC)~\cite{Boer:2011fh,Accardi:2012qut,Aschenauer:2017jsk,AbdulKhalek:2021gbh} in the US, LHeC/FCC-he~\cite{Agostini:2020fmq} at CERN and EicC in China~\cite{Anderle:2021wcy}. Thanks to the high energies and luminosities available at these future facilities, it will be possible to perform multi dimensional ``proton imaging'' and accurately determine the hadron structure not only as a function of momentum fraction $x$, but also including the geometric profile and intrinsic transverse momenta. 

The color charge correlators provide fundamental information about the proton, and these correlators can be measured through various exclusive and inclusive processes at the EIC. For example, 
at leading order the light-cone gauge correlator is related to the average quark transverse momentum vector
and to the Sivers asymmetry~\cite{Burkardt:2003yg}. Furthermore,
in the mixed transverse momentum -- transverse coordinate space representation, the color charge correlator can be related to the  Wigner distribution~\cite{Lorce:2011kd,Belitsky:2003nz,Ji:2003ak}, which is the most fundamental object describing the proton structure, and to various other generalized parton distribution functions. This detailed information about the partonic structure of the proton can be accessed experimentally   e.g.\ in dijet production or in vector meson -- lepton azimuthal correlations in DIS, as recently argued in Refs.~\cite{Hatta:2016dxp,Hatta:2017cte,Mantysaari:2020lhf,Mantysaari:2019csc}.

The first goal of this Letter is to present novel numerical results for the two-point color charge correlator in the proton at moderately small longitudinal momentum fraction $x$. Our numerical implementation is based on the recent computation presented in Ref.~\cite{Dumitru:2020gla}, where the two-point color charge correlator 
was computed within the framework of light-cone perturbation theory including the  next-to-leading order (NLO) corrections due to emission or exchange of a perturbative gluon\footnote{The two-point color charge correlator at leading order was first computed in Refs.~\cite{Dumitru:2020fdh,Dumitru:2018vpr}.}. 

Our results show non-trivial features in the proton color field at moderate $x$. For example, one may naively assume the impact parameter dependence of these color charge correlations to follow the proton geometry, i.e.\ to be proportional to the transverse shape function of the proton. We, however, find that this geometric picture may only apply at large impact parameters; while the nature of color charge correlations near the center of the proton is much more intricate. In particular, our results show that the color charge correlation function depends not
only on the impact parameter, but also on the relative
transverse momentum of the two gluon probes and on
the angle made by these two vectors. This is in contrast to
the frequently used, intuitive, McLerran-Venugopalan (MV) model~\cite{McLerran:1993ni, McLerran:1993ka}.
Also, for some kinematic configurations, the correlator may in fact display negative ("repulsive") correlations.
The need for non-trivial correlations among ``hot spots'' in
models for proton-proton scattering at high energies has been pointed
out previously~\cite{Albacete:2016pmp, Albacete:2016gxu,
  Albacete:2017ajt}.

Our second goal here is to study the evolution of the dipole-proton scattering amplitude at moderately small $x$ values (say $x\sim 0.1,\ldots,0.01$), and to present initial conditions for high-energy quantum chromodynamics (QCD) evolution to yet smaller $x$.
At high energies, where one is sensitive to the target structure at small $x$, the parton densities are so large~\cite{Abramowicz:2015mha} that individual quarks and gluons are not convenient degrees of freedom anymore. Instead, one considers eikonal interactions of the projectile with the effectively quasi-classical color field generated by the large $x$ partons in the target. 

To describe QCD in this regime of high parton densities, an effective theory of QCD, the Color Glass Condensate (CGC) has been developed~\cite{Iancu:2003xm,Kovner:2005pe,Gelis:2010nm,Blaizot:2016qgz}. In this approach, perturbative evolution equations such as the Balitsky-Kovchegov (BK) equation~\cite{Kovchegov:1999yj,Balitsky:1995ub} describe the energy (or $x$) evolution of scattering amplitudes. The initial condition for this evolution encodes non-perturbative information about the proton structure, and has been previously determined by fitting the HERA structure function data~\cite{Albacete:2010sy,Lappi:2013zma,Beuf:2020dxl}. 

In fact, the BK equation
in its standard formulation evolves the wave function of the
projectile, and the evolution ``time'' is given by the rapidity
of the {\em projectile}~\cite{Beuf:2014uia,
  Ducloue:2019ezk, Boussarie:2020fpb}. Duclou\'e {\em et al.} have
reformulated~\cite{Ducloue:2019ezk} BK evolution at NLO in terms of
the target rapidity, which
is the rapidity related to $x$. The resulting evolution equation is non-local in rapidity, which  underscores the
importance of $x$-dependent ``initial conditions'' for the dipole
scattering amplitude as computed here. We emphasize that such an initial condition is a necessary ingredient for all phenomenological applications of the CGC framework; these applications are currently reaching NLO accuracy~\cite{Lappi:2020ufv,Beuf:2020dxl,Ducloue:2017ftk,Lappi:2015fma,Lappi:2016fmu,Hanninen:2017ddy,Lappi:2016oup,Boussarie:2016bkq,Escobedo:2019bxn,Beuf:2016wdz,Altinoluk:2011qy,Altinoluk:2014eka,Roy:2019hwr}, and this calls for a rigorous NLO level calculation of the initial condition. We also note that the calculation of the color charge correlator at NLO accuracy is required in order to determine the dependence on $x$, which
enters as a cutoff on the gluon longitudinal momentum fraction, as discussed below.

%-------------------------------------------------------------
\section{Color charge correlator}

The central object of consideration in this paper is the two-point
color charge correlator\footnote{Note that here we do not extract from
  $G_2$ the normalization factor $C(N)=\frac{1}{2}$ of the generators
  of the fundamental representation, as was done in
  Ref.~\cite{Dumitru:2018vpr,Dumitru:2020fdh}.}
\begin{equation}
\label{eq:g2def}
\langle \rho^a(\vec q_1)\, \rho^b(\vec q_2) \rangle \equiv \delta ^{ab}\, g^2
G_2(\vec q_1,\vec q_2),
\end{equation}
where $g = \sqrt{4\pi\as}$ is the strong coupling constant and $a,b$ are the external gluon colors.
The notation $\langle\cdots\rangle$ denotes an expectation value between proton
states, $\langle K|$ and $|P\rangle$, stripped of the delta-functions for
conservation of transverse and light-cone momentum\footnote{We use the light cone  coordinates $ (P^{+},P^{-},\vec P)$,  where the  notation $\vec P$ denotes  two-dimensional transverse vector.}
\begin{multline}
    \langle K|\,
\rho^a(\vec q_1)\, \rho^b(\vec q_2)\, |P\rangle = 16\pi^3\, P^+ \\
\times 
\delta(P^+ - K^+)\, \delta(\vec P - \vec K - \vec q_1 - \vec q_2)\,
\langle \rho^a(\vec q_1)\, \rho^b(\vec q_2) \rangle.
\end{multline}
Furthermore, $\rho^a(\vec q)=\rho^a_\mathrm{qu}(\vec q)+\rho^a_\mathrm{gl}(\vec q)$ denotes
the operator which sums up the color charge densities of quarks (qu) and gluons (gl) with
$q^+>0$; explicit expressions in terms of quark and gluon creation and annihilation
operators are given in Ref.~\cite{Dumitru:2020gla}.

The insertion of
the charge operators $\rho^a(\vec q_1),\rho^b(\vec q_2)$ between the incoming and scattered proton states
corresponds to the attachment of two static gluon probes (with
amputated propagators) to the color charges in the proton, in all
possible ways. The complete set of diagrams for $G_2$ at NLO and
their explicit expressions are given in Ref.~\cite{Dumitru:2020gla}. The two static
gluons that probe the proton structure carry transverse momenta $\vec q_1$ and $\vec q_2$, and the total momentum transfer to the proton reads
\begin{equation}
    \vec P - \vec K = \vec q_1  + \vec q_2.
\end{equation}
In subsequent expressions, such as Eq.~(\ref{eq:G2_q12_b}), we choose $\vec P=0$ for the incoming proton\footnote{Note that the computed correlator is invariant under the transverse Galilean transformations, see the  discussion in Ref.~\cite{Dumitru:2020gla}.}.
\begin{figure}[tb]
\includegraphics[width=0.20\textwidth]{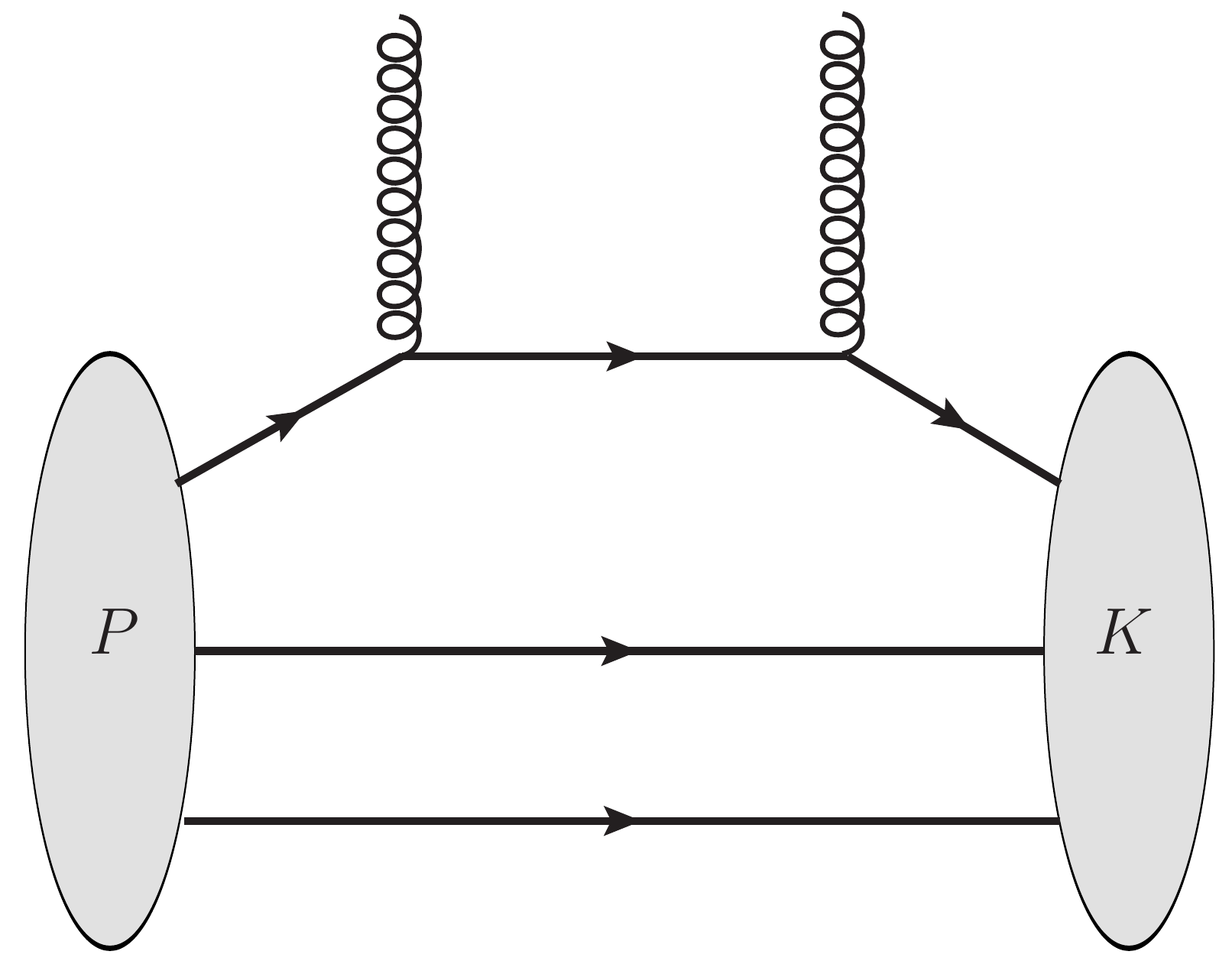}
\includegraphics[width=0.20\textwidth]{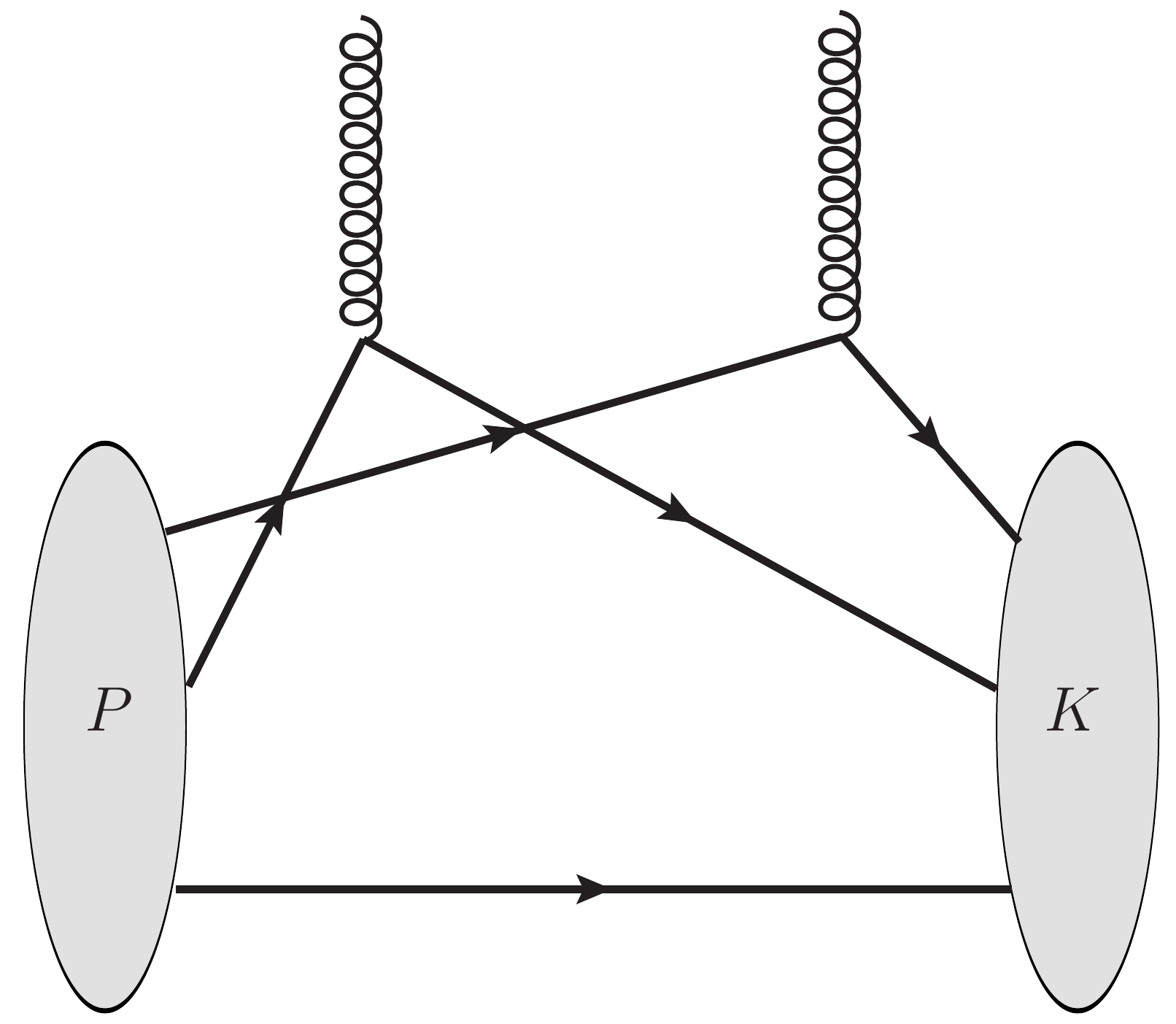}
\includegraphics[width=0.20\textwidth]{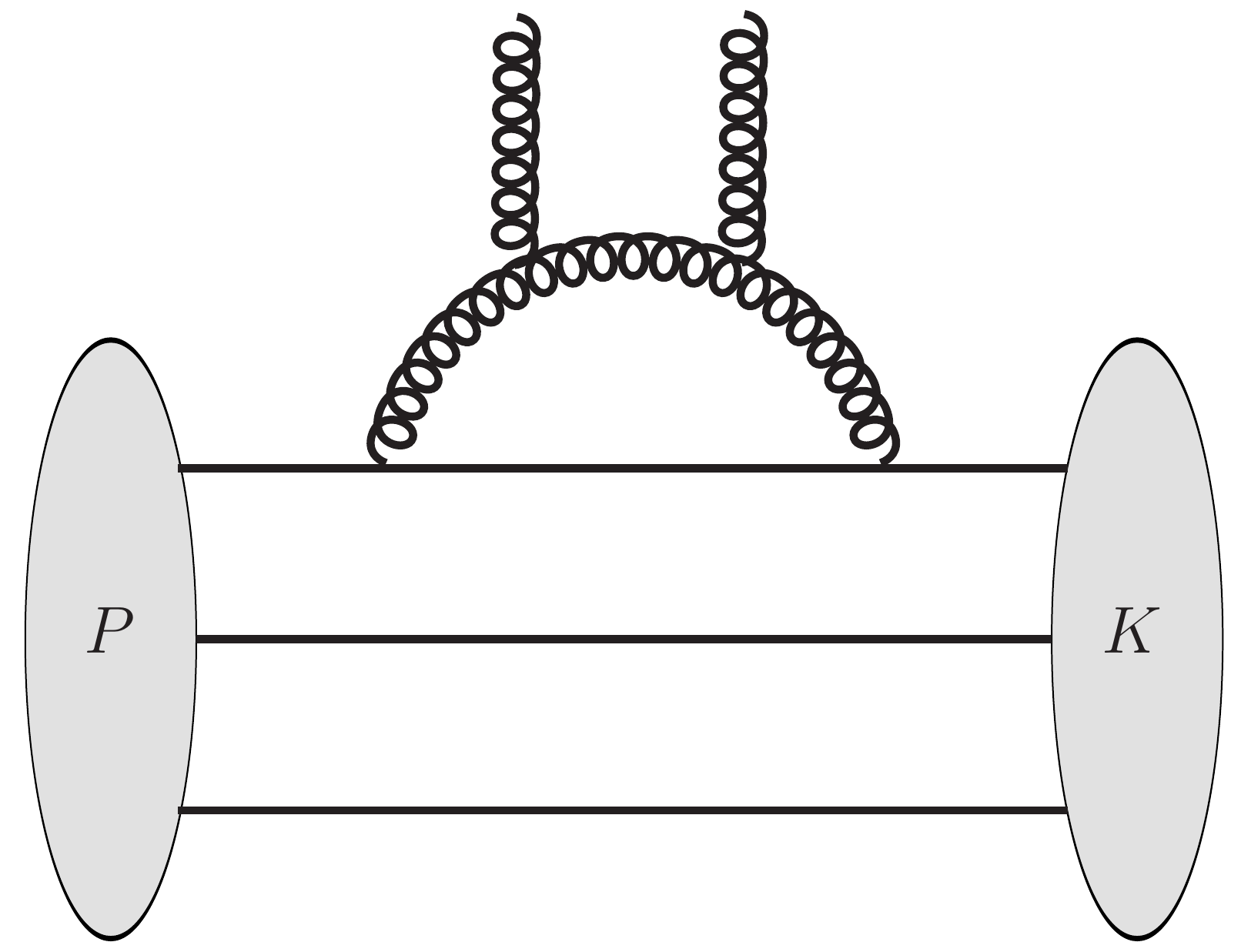}
\includegraphics[width=0.20\textwidth]{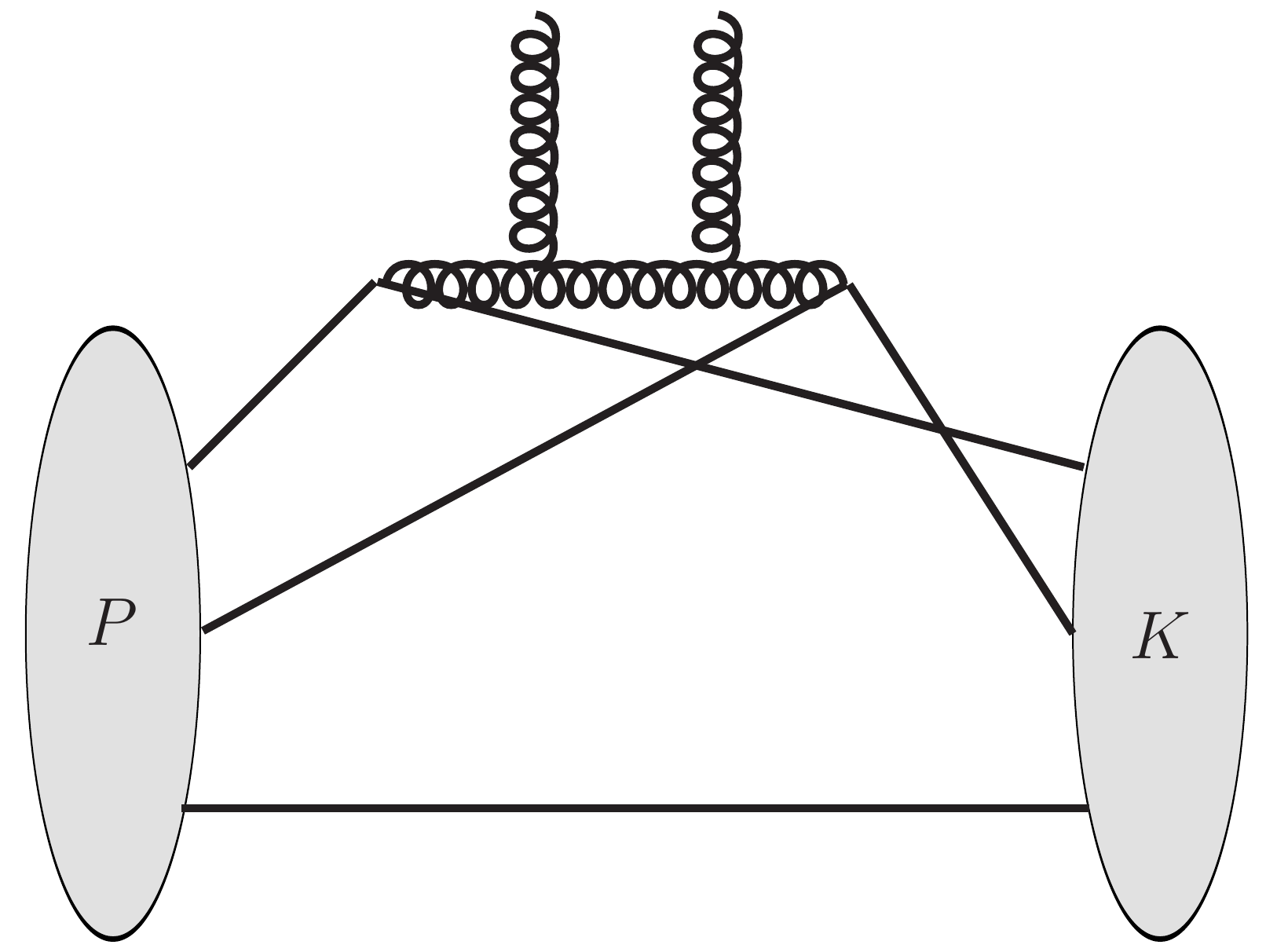}
\caption{Examples for handbag (left) and cat's ears (right) diagrams at LO (top) and at NLO (bottom).}
\label{fig:Handbag-CatsEars}
\end{figure}
Figure~\ref{fig:Handbag-CatsEars} shows examples for ``handbag'' and
``cat's ears'' diagrams at leading order (LO) $g^2$ and at NLO $g^4$ in the strong coupling constant. The former are represented at LO by
one-body operators so that the entire momentum transfer $\vec K$ to
the proton flows into a single valence quark line. Therefore, for
$\vec q_1\to-\vec q_2$ (i.e.\ $\vec K \to 0$) the LO handbag diagrams
approach the normalization integral in Eq.~(\ref{eq:Norm}), such that there is
maximal wave function overlap. At leading order, the handbag diagram is
proportional to the electromagnetic form factor, i.e.\ to the
distribution $\langle\rho(\vec q)\rangle$ of electric charge in the
proton\footnote{See, for example, Eqs.~(9,10) in~\cite{Dumitru:2019qec}.}.

For large momentum transfer, on the
other hand, wave function overlap in the handbag diagram is highly suppressed. At
large $\vec K^2$, the wave function overlap is much greater for the
cat's ears diagram because the momentum transfer is shared by two (or
even three, at NLO) valence quarks. Since $\vec K$ is the Fourier
conjugate to the two-dimensional (2D) transverse coordinate vector (impact parameter)
$\vec b$, it follows that color charge correlators near the center of
the proton are dominated by diagrams where the momentum transfer is
shared by the valence quarks~\cite{Dumitru:2020fdh,Dumitru:2019qec}.

The gluon emission and exchange diagrams exhibit ultraviolet
divergences when the gluon transverse momentum $\vec k_g \to
\infty$. These cancel in the sum of diagrams~\cite{Dumitru:2020gla} so
that $G_2$ is renormalization scale independent. They also exhibit
soft and collinear divergences, which are regularized by introducing a
light-cone momentum cutoff $x$, and a collinear cutoff $m$  in
the light-cone energy denominators, respectively following the notation of Ref.~\cite{Dumitru:2020gla}\footnote{The
  dependence on $x$ and $m$ is left implicit, we do not list
  these cutoffs as arguments of $G_2$.}.  The dependence of the
correlator on $x$ and $m$ will be explored numerically below.

The Fourier transform w.r.t.\ the total momentum transfer to the
proton gives the color charge correlator as a function of impact
parameter $\vec b$ and the relative transverse momentum $\vec q_{12} =
\vec q_1 - \vec q_2$ of the probes
\be  \label{eq:G2_q12_b}
G_2(\vec q_{12},\vec b) = \! \! \int \! \frac{\der^2 \vec K}{(2\pi)^2} e^{-i\vec b\cdot \vec K}\,
\!G_2\!\left(\frac{\vec q_{12}-\vec K}{2}, -\frac{\vec q_{12}+\vec
  K}{2} \right).
\ee
The vector $\vec q_{12}$ is Fourier conjugate to the transverse distance
$\vec r$ between the two gluons. For $\vec q_{12}=0$, the integral of $G_2$ over
the transverse impact parameter plane vanishes,
\be
\int \der^2 \vec b \,\, G_2(\vec q_{12}=0,\vec b) = 0.
\label{eq:G2-sum_rule}
\ee
This is due to the fact that $G_2(\vec q_1,\vec q_2)$ satisfies a Ward
identity and vanishes when either $\vec q_i\to0$~\cite{Bartels:1999aw,
  Ewerz:2001fb, Dumitru:2020fdh, Dumitru:2020gla}.

From the color charge correlator we can obtain the eikonal dipole scattering
amplitude $N(\vec r,\vec b)$ in the two-gluon exchange approximation
as follows~\cite{Dumitru:2018vpr}
\begin{multline}
  N(\vec r,\vec b) = -g^4 C_F
  \int \frac{\der^2 \vec K \der^2 \vec q}{(2\pi)^4}
  \frac{\cos\left(\Vec b \cdot \vec K\right)}{(\vec q - \frac{1}{2}\vec K)^2\,\,
    (\vec q + \frac{1}{2} \vec K)^2} \\
  \times \left( \cos(\vec r \cdot \vec q) -
  \cos\left(\frac{\vec r \cdot \vec K}{2} \right) \!\! \right)
  \! G_2\left(\vec q -\frac{1}{2}\vec K, -\vec q - \frac{1}{2} \vec K\right). 
\end{multline}
This expression applies in the regime of weak scattering, $N(\vec
r,\vec b) \ll 1$ since it does not resum the Glauber-Mueller multiple
scattering series. To perform such resummation, the color charge
correlator computed in Ref.~\cite{Dumitru:2020fdh} would have to be
transformed from light cone to covariant gauge\footnote{For a large
  nucleus where the color charge correlator of the MV model applies
  this has been done in Ref.~\cite{Kovchegov:1996ty}.}.
Also, we stress that the LO contribution to $N(\vec r,\vec b)$ is proportional
to $\as^2$ while the NLO correction is proportional
to $\as^3$ (times a logarithm of the minimal light-cone momentum fraction $x$
of the gluon, in a leading logarithmic approximation).

It will be instructive to contrast our result for $G_2(\vec q_1,\vec
q_2)$ to the color charge correlator of the McLerran-Venugopalan (MV)
model~\cite{McLerran:1993ni, McLerran:1993ka}. In transverse momentum space,
\be
\langle \rho^a(\vec q_1)\, \rho^b(\vec q_2) \rangle = \delta^{ab}\,
g^2\, \mu^2\, (2\pi)^2\delta(\vec q_1 + \vec q_2).
\ee
Here, $g^2 \mu^2>0$ has the interpretation of the mean square color charge
per unit area. In impact parameter space this corresponds to $G_2(\vec
b, \vec q_{12}) \to \mu^2$, i.e.\ to a translationally invariant
correlator which is independent of the relative transverse momentum
(and its azimuthal orientation relative to $\vec b$). In applications,
where a non-trivial transverse profile functions is required, one
commonly replaces in impact parameter space
$\mu^2 \to \mu^2(b)$, in which $\mu^2(b)\sim T_p(b)$ is
proportional to the proton shape function~\cite{Schenke:2012hg,Schenke:2012wb,Iancu:2017fzn} (see also Refs.~\cite{Kovner:2018fxj,Mantysaari:2020lhf,Mantysaari:2019jhh,Mantysaari:2019csc,Mantysaari:2018zdd,Mantysaari:2016ykx,Mantysaari:2016jaz,Demirci:2021kya} for phenomenological applications).  Such a dependence of the correlator
on impact parameter does not satisfy the sum
rule Eq.~(\ref{eq:G2-sum_rule}) without an explicit confinement scale regulator. Furthermore, the MV-model correlator does
not exhibit the soft and collinear divergences of the NLO correlator
$G_2$.

%-------------------------------------------------------------

\section{The proton state on the light front}

\begin{figure}[htb]
\includegraphics[width=0.5\textwidth]{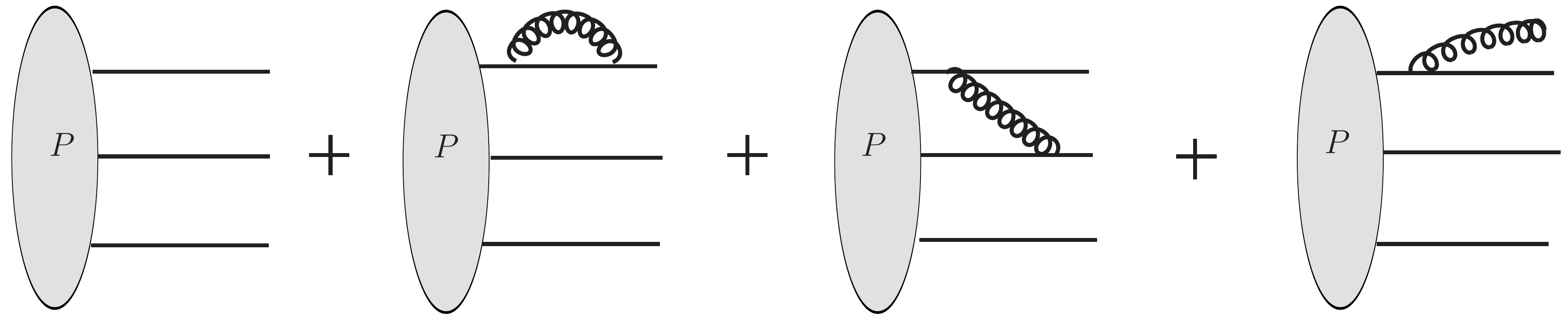}
\caption{Diagrammatic representation of the proton state $|P\rangle$
  up to NLO. }
\label{fig:Proton-state-NLO}
\end{figure}
At LO the proton state on the light cone is approximated by a
three-quark Fock state with the wave function $\Psi_{qqq}(x_1,\vec
k_1; x_2,\vec k_2; x_3,\vec k_3)$~\cite{Brodsky:1997de,Schlumpf:1992vq,Brodsky:1994fz}. Here, $\vec
k_i$ are the transverse momenta of the on-shell quarks in the
transverse rest frame of the proton. At NLO, there are ${\cal O}(g)$
corrections due to the emission of a gluon by one of the quarks as
well as ${\cal O}(g^2)$ virtual corrections due to the exchange of a
gluon by a quark either with itself or with another quark. This is
shown in Fig.~\ref{fig:Proton-state-NLO}; explicit expressions are
given in Ref.~\cite{Dumitru:2020gla}.

The valence quark wave function $\Psi_{qqq}$ is gauge invariant and universal,
process independent. It encodes the non-perturbative structure of the proton
at low transverse momentum scales, and light cone momentum fraction in the
valence quark regime. For our numerical estimates here, we employ a simple model
due to Brodsky and Schlumpf~\cite{Brodsky:1994fz}
\begin{equation}
  \Psi_{qqq} = N_{\mathrm{HO}}
  \sqrt{x_1 x_2 x_3}
  \prod_{i=1}^3 \exp\left( -\frac{(\vec k_i^2+M^2)/x_i}{2\beta^2} \right).
  \label{eq:Psiqqq_HO}
\end{equation}
This ``harmonic oscillator'' wave function is our default choice. For
some observables we have also checked the power-law wave function
\begin{equation}
  \Psi_{qqq} = N_{\mathrm{p}}
  \sqrt{x_1 x_2 x_3} \left(1+\sum_{i=1}^3
  \frac{\vec k_i^2+\tilde M^2}{x_i\tilde \beta^2}\right)^{-p}.
  \label{eq:Psiqqq_P}
\end{equation}
The non-perturbative parameters $M=0.26\,\gev$ and $\beta=0.55\,\gev$, introduced in
Eq.~\eqref{eq:Psiqqq_HO}, have been tuned in
Ref.~\cite{Brodsky:1994fz} to reproduce the radius, the anomalous
magnetic moment, and the axial coupling of the proton and the neutron. Similarly parameters for the power-law wave function \eqref{eq:Psiqqq_P} are available in Ref.~\cite{Brodsky:1994fz}. Note that because of
the constraints that $x_1+x_2+x_3=1$ and $\vec k_1+\vec k_2+\vec
k_3=0$, the proton state also encodes longitudinal and transverse
momentum correlations among the quarks.

We have observed rather small differences in the kinematic regimes
covered by the figures below. We therefore do not show explicitly the results corresponding to the
power law valence quark wave function. The normalization factor of $\Psi_{qqq}$ is obtained from the requirement that the states are normalized as
\bea
\langle K | P\rangle &=& 16\pi^3\, P^+\, \delta(P^+ - K^+)\,
\delta(\vec P - \vec K) \nonumber\\ &\to&
\frac{1}{2}\int[\der x_i] \int[\der^2 \vec k_i]\,\, |\Psi_{qqq}(x_i,\vec k_i)|^2 = 1,
\label{eq:Norm}
\eea
where the phase space factors are defined as:
\bea
[\der x_i] &\equiv& \frac{\der x_1 \der x_2 \der x_3}{8 x_1 x_2 x_3}\, \delta(1-x_1-x_2-x_3), \\
\left[\der^2 \vec k_i\right] &\equiv&
\frac{\der^2 \vec k_1 \der^2 \vec k_2 \der^2 \vec k_3}{(2\pi)^6}\, \delta(\vec k_1+\vec k_2+\vec k_3).
\eea

Other models for the valence quark wave functions include those of
Refs.~\cite{Frank:1995pv,Miller:2002ig,Pasquini:2007iz,Pasquini:2009bv,
  Lorce:2011dv}. Valence quark wave functions have also been obtained
numerically by solving various model Hamiltonians~\cite{Zhao:2020gtf,
  Xu:2020xbt, Mondal:2020mpv}.

%-------------------------------------------------------------

\section{Results and discussion}

For all numerical results we use fixed coupling\footnote{The coupling does not run as the perturbative one gluon emission corrections are  $\mathcal{O}(\as)$, see discussion in Ref.~\cite{Dumitru:2020gla}} $\as=0.2$. We note that NLO corrections increase
relative to the LO results in proportion to $\as$. Also, unless
mentioned otherwise, our default choice for the collinear cutoff is
$m=0.2$~GeV. We show results at different $x$, which is the lower cutoff for the emitted gluon longitudinal momentum in the NLO diagrams. We also use $x$ as a lower limit in the integrations over the valence quark momentum fractions $x_i$.

\subsection{Color charge density correlator}

\begin{figure}[tb]
\includegraphics[width=0.5\textwidth]{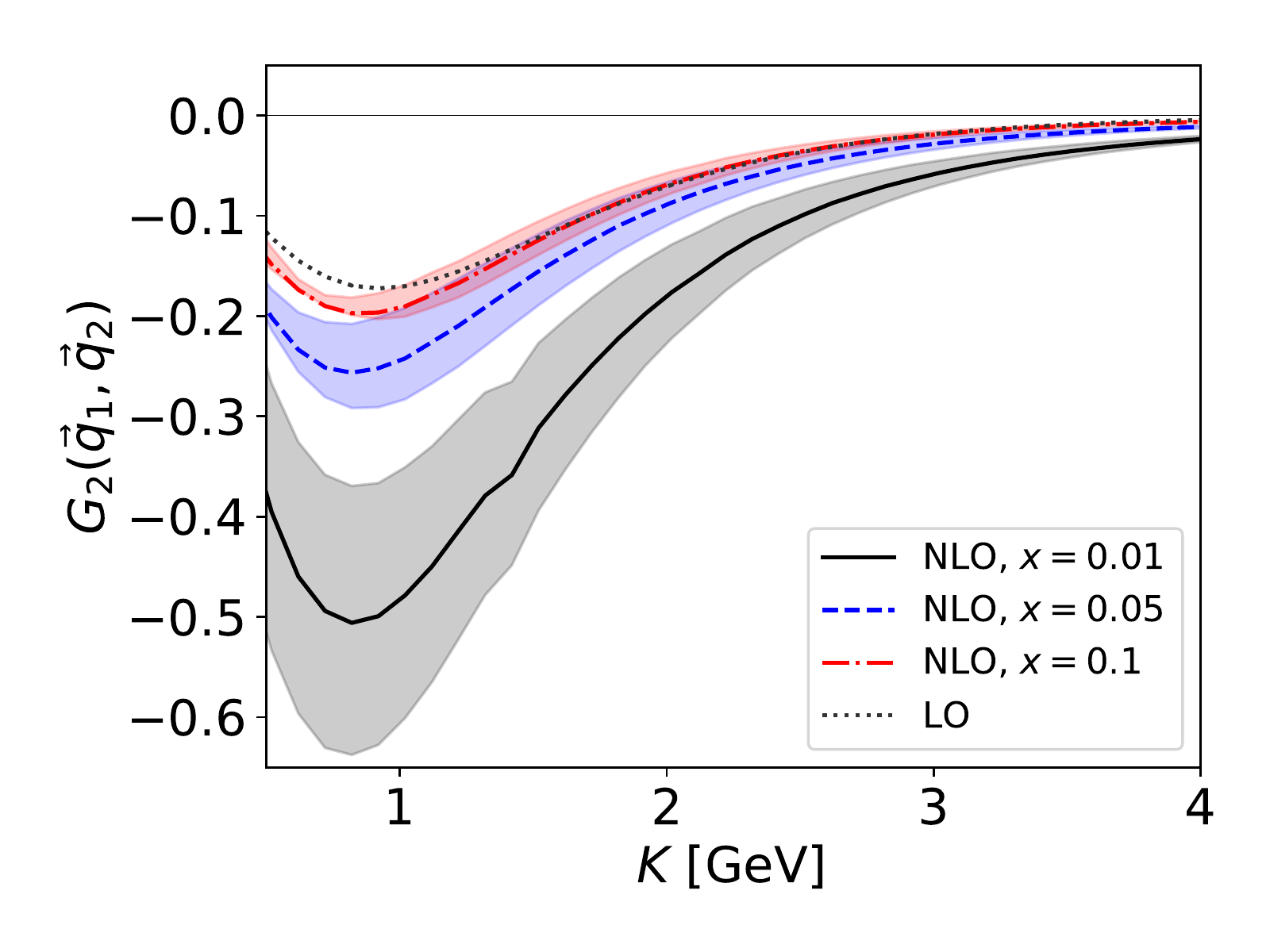}
\caption{Color charge correlator in momentum space at different $x$ for the
configuration where $\vec q_1=\vec q_2=(K/2,0)$ are parallel. The results are shown in the perturbative $K>0.5\,\gev$ region.
The bands correspond to the variation of the collinear cutoff in $m=0.1\,\gev \dots 0.4\, \gev$. 
}
\label{fig:momspace_xdep}
\end{figure}
To study the magnitude of the NLO correction, we evaluate the two
point function $G_2(\vec q_1,\vec q_2)$ in momentum space.  We first
choose a configuration where the two probe gluons carry the same
momentum, $\vec q_1=\vec q_2=(K /2,0)$. Note that here, and in the following discussion, we use the notation $K = |\vec K|$ for the length of all transverse vectors. The color charge correlator
as a function of the transverse momentum transfer
$K$ is shown in Fig.~\ref{fig:momspace_xdep}. The
leading order result is compared to the full NLO result (which also
includes the leading order $\sim g^2$ contribution) obtained with
three different values for the momentum cutoff $x$.  As expected, the
one gluon emission correction grows rapidly with decreasing $x$, and
completely dominates at $x=0.01$. At greater $x= 0.05$ the
NLO correction is comparable to the leading order contribution, to finally
become a small correction at $x=0.1$. 
The fact that the perturbative correction is numerically small at $x=0.1$ represents
a non-trivial consistency check of the "expansion" about the LO three-quark state.
We also observe that the dependence on the collinear cutoff $m$ is moderate, as is the case in all results shown in this Section.

We have mentioned above that at leading order, the configuration with
approximately equal momenta is dominated by the cat's ears diagram,
which corresponds to the matrix element of a two-body operator and
gives a negative contribution to $G_2$. The figure shows that the
correlation function for $\vec q_1 = \vec q_2$ remains negative,
i.e.\ dominated by two-body correlations, even in the regime of small
$x$ where the NLO correction is greater than the LO contribution.

\begin{figure}[tb]
\includegraphics[width=0.5\textwidth]{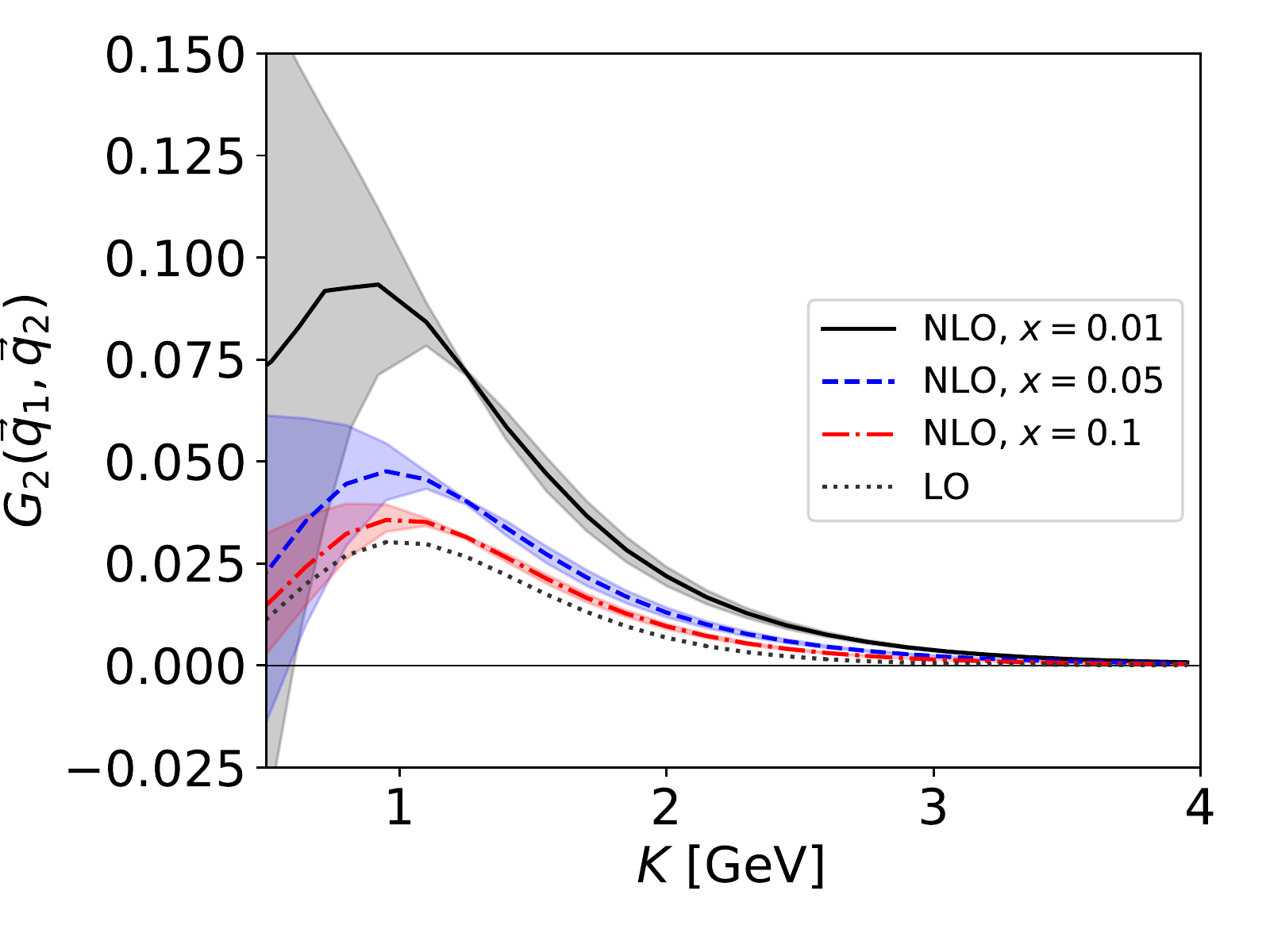}
\caption{Color charge correlator in momentum space at different $x$
  for the configuration where the probe momenta are perpendicular:
  $\vec q_1 = (K/\sqrt{2},0)$ and $\vec q_2 = (0,K/\sqrt{2})$,
  so $q_{12} = K$. The results are shown in the perturbative $K>0.5\,\gev$ region. The bands correspond to the variation of the collinear cutoff in $m=0.1\,\gev \dots 0.4\, \gev$.
  }
\label{fig:momspace_xdep-perp}
\end{figure}
In Fig.~\ref{fig:momspace_xdep-perp} we show $G_2(\vec q_1,\vec q_2)$
for perpendicular momenta where contributions of hand-bag type are
less suppressed. Indeed, the correlator is now positive,
although this contribution is generically smaller in magnitude
than the negative
contribution from the cat's ears type diagrams shown in the previous
Fig.~\ref{fig:momspace_xdep}. 
The NLO correction amounts to
stronger positive correlations at all momentum transfers $K$. At small $K\lesssim 1\,\gev$ the results depend strongly on the value chosen for the collinear cutoff $m$.
Of course, at small $K$ {\em and}
small $q_{12}$ the perturbative computation
of the color charge correlator in terms of two gluon exchange should
be interpreted with caution.

\begin{figure}[tb]
\includegraphics[width=0.5\textwidth]{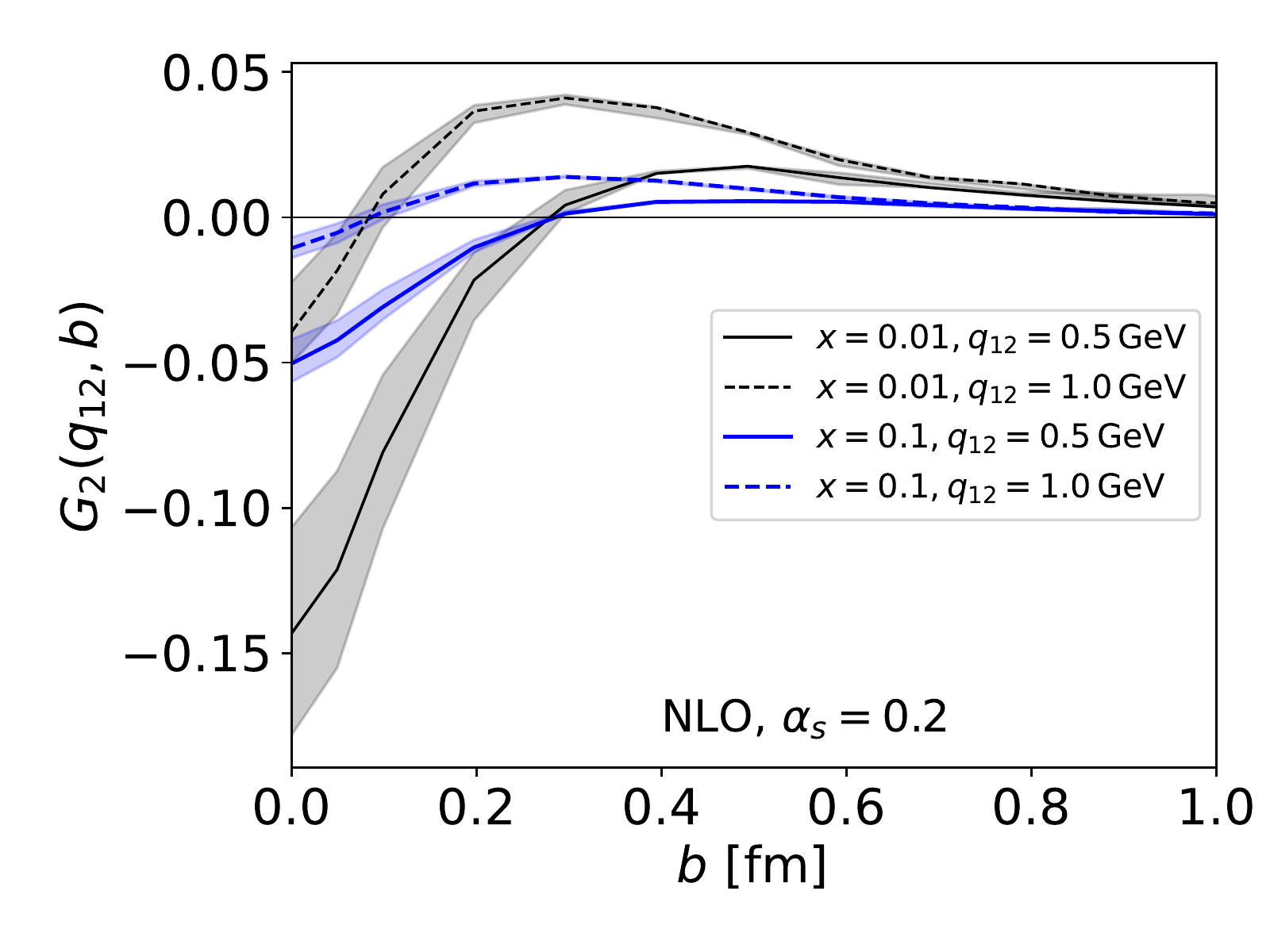}
  \caption{Color charge correlator $G_2(\vec q_{12}, \vec b)$ as a function of impact
    parameter for various relative transverse momenta $q_{12}$ averaged over the angle between $\vec q_{12}$ and $\vec b$. The bands correspond to the variation of the collinear cutoff in $m=0.1\,\gev \dots 0.4\, \gev$.
  }
\label{fig:mixed_space_finite_q12}
\end{figure}

We now proceed to show the color charge correlator $G_2(\vec q_{12},
\vec b)$ in mixed transverse momentum - transverse coordinate space,
c.f.\ Fig.~\ref{fig:mixed_space_finite_q12}.
Near the center of the proton at small $b$ we find that $G_2<0$;
``repulsive'' two-body correlations dominate here. Also, comparing
$x=0.1$ to $x=0.01$ we note that the NLO correction
mainly affects $G_2$ at small $b$ to strongly boost these negative two-body
correlations, more so for smaller $q_{12}$. However, hand-bag type contributions become more
prominent with increasing $q_{12}$ or $b$. At small $x=0.01$ and large
$q_{12} = 1$~GeV, we observe that significant positive color charge
correlations emerge for impact parameters $b \agt
0.2$~fm. Generically, the large-$b$ tails of the two-point
correlator $G_2$ exhibit a fall-off that resembles a transverse
profile function.
Here, the one gluon emission NLO
correction is in line with the simple intuitive picture whereby it increases the mean square color charge density.
However, there is a large
negative correction near the center of the proton.

\begin{figure}[tb]
\includegraphics[width=0.5\textwidth]{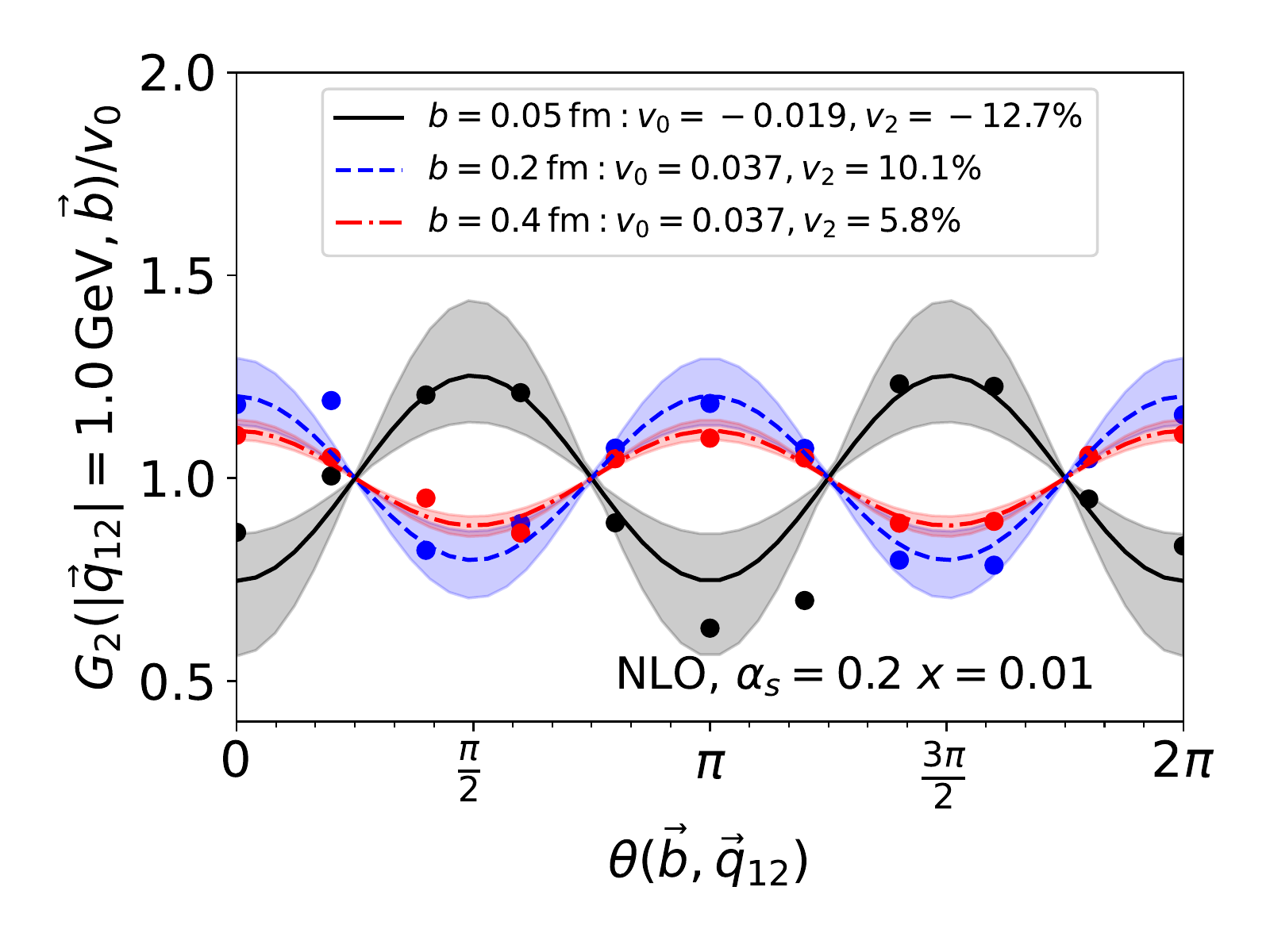}
  \caption{$G_2(\vec q_{12}, \vec b)$ as a function of the azimuthal
    angle made by $\vec q_{12}$ and $\vec b$ at three different impact
    parameters. Curves show fits of the form $v_0 (1+ 2v_2\cos 2\theta$). The bands correspond to the variation of the collinear cutoff in $m=0.1\,\gev \dots 0.4\, \gev$. 
    }
\label{fig:G2-angular}
\end{figure}
The charge correlator in the MV model, even in its modified version
with a non-trivial shape function, is independent of the relative
transverse momentum $\vec q_{12}$ and its angular orientation relative
to the impact parameter vector $\vec b$. We have already documented
above the dependence of our result for $G_2$ on the magnitude of $\vec
q_{12}$. Fig.~\ref{fig:G2-angular} shows that it depends also on the
relative azimuthal orientation $\theta$ of these vectors. We observe a
$\sim \cos 2(\theta-\varphi)$ azimuthal anisotropy in $G_2$ normalized
by its angular average. The phase shift is
$\varphi=0$ in the kinematic regime where the correlator is positive,
and $\varphi=\pi/2$ in the regime dominated by ``repulsive''
correlations where $G_2(\vec q_{12},\vec b)<0$.

%-------------------------------------------------------
\subsection{Dipole scattering amplitude $N(\vec r,\vec b)$}

In this section, we document the "evolution" of the dipole scattering
amplitude from $x=0.1$ to $x=0.01$, and its angular dependence. We have
not attempted to tune the coupling constant 
or the quark mass collinear cutoff
to data from inclusive DIS or exclusive $J/\Psi$ production;
detailed studies of phenomenological predictions are left for the future. 

\begin{figure}[tb]
\includegraphics[width=0.5\textwidth]{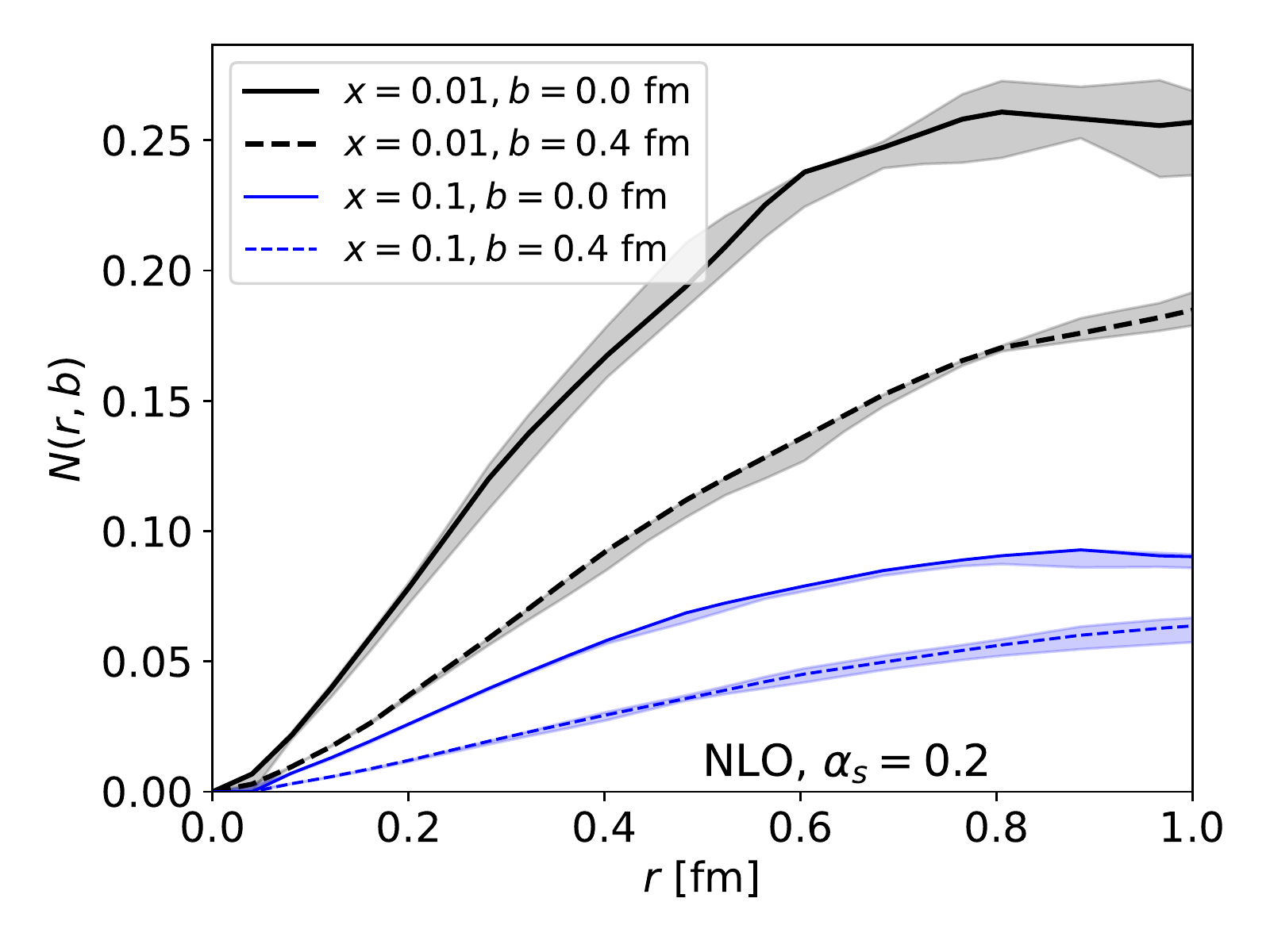}
\caption{Dipole scattering amplitude at NLO  as a
  function of dipole size $r$ at impact parameters $b=0$ and at $b=0.4\,\mathrm{fm}$ at two different $x$ averaged over the angle between $\vec r$ and $\vec b$. The bands correspond to the variation of the collinear cutoff in $m=0.1\,\gev \dots 0.4\, \gev$.}
\label{fig:dipole_lo_vs_nlo}
\end{figure}
In Fig.~\ref{fig:dipole_lo_vs_nlo} we show the evolution of the dipole
scattering amplitude from $x=0.1$ to $x=0.01$, at two different impact
parameters. We observe much stronger scattering at higher energy
(lower $x$) by a factor of $\approx 3$, even though $\alpha_s\log
\frac{0.1}{0.01}=0.46$ is not a big number. However, there are many
diagrams for NLO corrections. The obtained dipole amplitude $N(\vec r,\vec b;x)$ can be directly used as an initial
condition for the (impact parameter dependent) BK evolution studied e.g. in Refs.~\cite{GolecBiernat:2003ym,Berger:2010sh,Berger:2011ew,Berger:2012wx,Cepila:2018faq, Bendova:2019psy,Cepila:2020xol,Bendova:2020hkp}.

We continue with an analysis of the azimuthal anisotropy of the dipole
scattering amplitude. If the color charge correlator is taken to be
isotropic and proportional to the shape function of the target then, in the two-gluon exchange approximation
at small $r$ and $b$,
\be \label{eq:N_MV}
N(\vec r, \vec b) = f(r,b)\, \left[ 1 + c\, (r b)^2 \cos 2\theta
  \right].
\ee
Here, the function $f(r,b)$ is independent of the angle $\theta$
between $\vec r$ and $\vec b$, 
and $c>0$ is a constant with mass dimension 4. See Ref.~\cite{Iancu:2017fzn} for a derivation of this
result\footnote{Also see Ref.~\cite{Levin:2011fb} and Sec.~6
  in~\cite{Kovner:2012jm}. Related considerations can be found in
  Refs.~\cite{Kopeliovich:2007fv, Kopeliovich:2008nx, Zhou:2016rnt,
    Hagiwara:2017ofm}.},
and Fig.~18 in Ref.~\cite{Mantysaari:2020lhf} for a nice graphical representation. 
Here, the angular dependence arises due to
the fact that under rotations of the dipole at fixed $\vec b$ its
endpoints probe different ``densities'' in the target, when its
transverse profile function is not constant. (The target is isotropic
only w.r.t.\ its center but is not invariant under rotations about the
displaced point $\vec b$). Eq.~(\ref{eq:N_MV}) predicts a $\sim \cos
2\theta$ azimuthal dependence with an amplitude proportional to the
squares of the size of the dipole and of the impact parameter.

\begin{figure}[tb]
\includegraphics[width=0.5\textwidth]{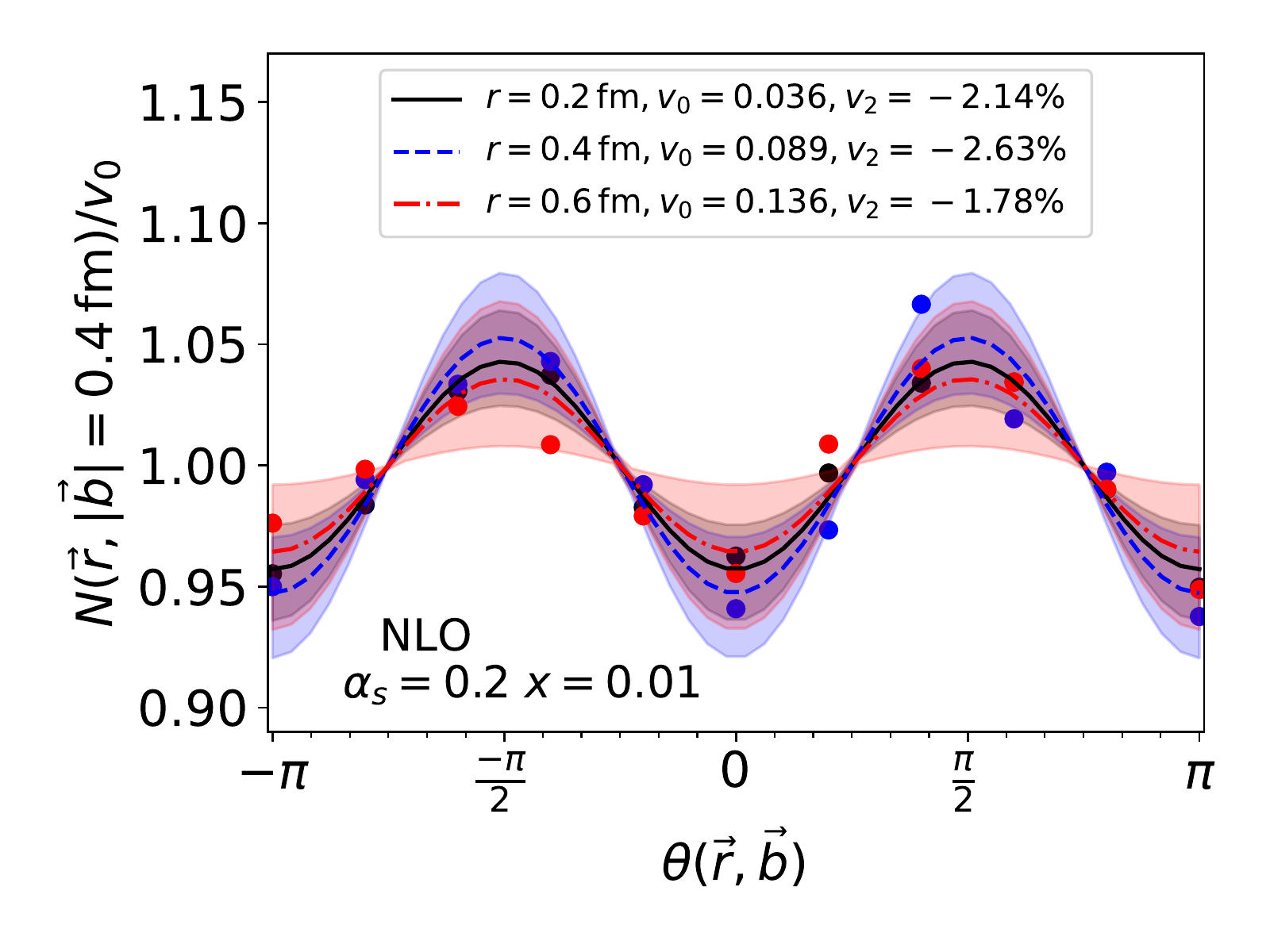}
\caption{Dipole scattering amplitude at NLO as a function of the angle
  between $\vec r$ and $\vec b$. Points show the actual result of a numerical calculation and lines are obtained by fitting a functional form $v_0(1+2v_2\cos2\theta)$.
 The bands correspond to the variation of the collinear cutoff in $m=0.1\,\gev \dots 0.4\, \gev$. 
 }
\label{fig:dipole_lo_vs_nlo_modulation}
\end{figure}
Figure~\ref{fig:dipole_lo_vs_nlo_modulation} shows our result for
$N(\vec r,\vec b)$ at fixed $b =0.4$~fm. It indeed displays a $\sim \cos
2\theta$ azimuthal dependence, but in contrast to MV model based calculations of Refs.~\cite{Iancu:2017fzn,Mantysaari:2019csc,Mantysaari:2020lhf} we find a phase shift of $\pi/2$, i.e. a negative $v_2 = \langle \cos 2\theta \rangle$ (or $c<0$ in Eq.~\eqref{eq:N_MV}).
Physically, our result corresponds to {\em stronger} scattering when the dipole is oriented perpendicular to the impact parameter. In the calculations based on the MV model one finds the opposite behavior where parallel alignment results in a greater scattering amplitude.
At the given impact parameter, we find that the amplitude of the
angular modulation is nearly constant for $r=0.2\dots0.6$~fm. This
indicates that the non-zero $\langle\cos 2\theta\rangle$ is not
entirely due to simply a non-trivial shape function of the target
proton. Indeed, the angular dependence of the color charge correlator
itself, which we have shown in the previous section, is also
important. The nearly
constant $v_2$ when $r=0.2\to0.6~\fm$ may support the ``domain'' picture introduced by Kovner and
Lublinsky~\cite{Kovner:2012jm}.

\begin{figure}[tb]
\includegraphics[width=0.5\textwidth]{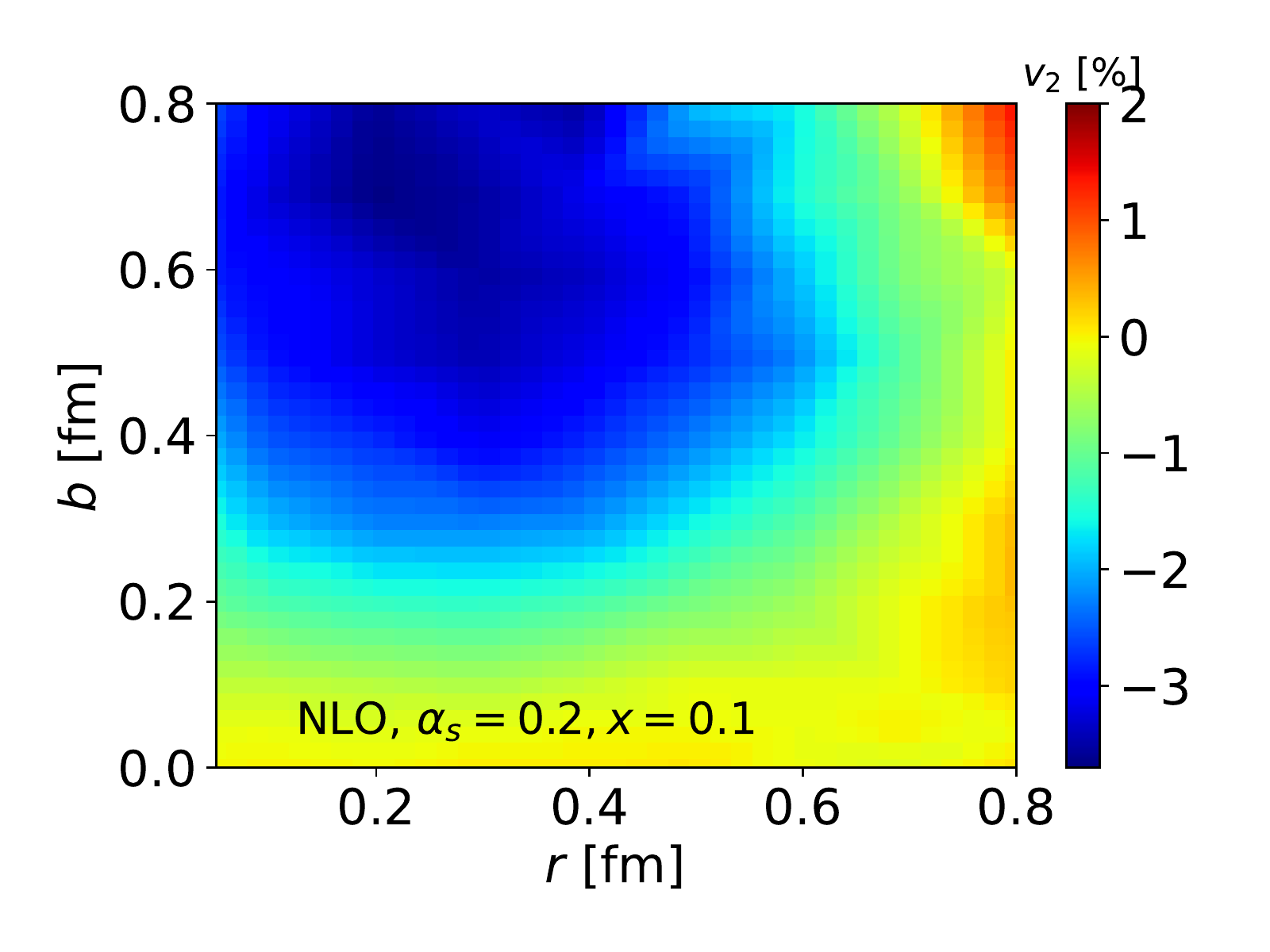}
\caption{$v_2=\langle\cos 2\theta\rangle$ from the dipole scattering
  amplitude at NLO in the impact parameter vs.\ dipole size plane  at $x=0.1$.
The smallest dipole size at the left edge of the plot is $r_\text{min}=0.05~\fm.$}
\label{fig:dipole_v2_heatmap_01}
\end{figure}

\begin{figure}[tb]
\includegraphics[width=0.5\textwidth]{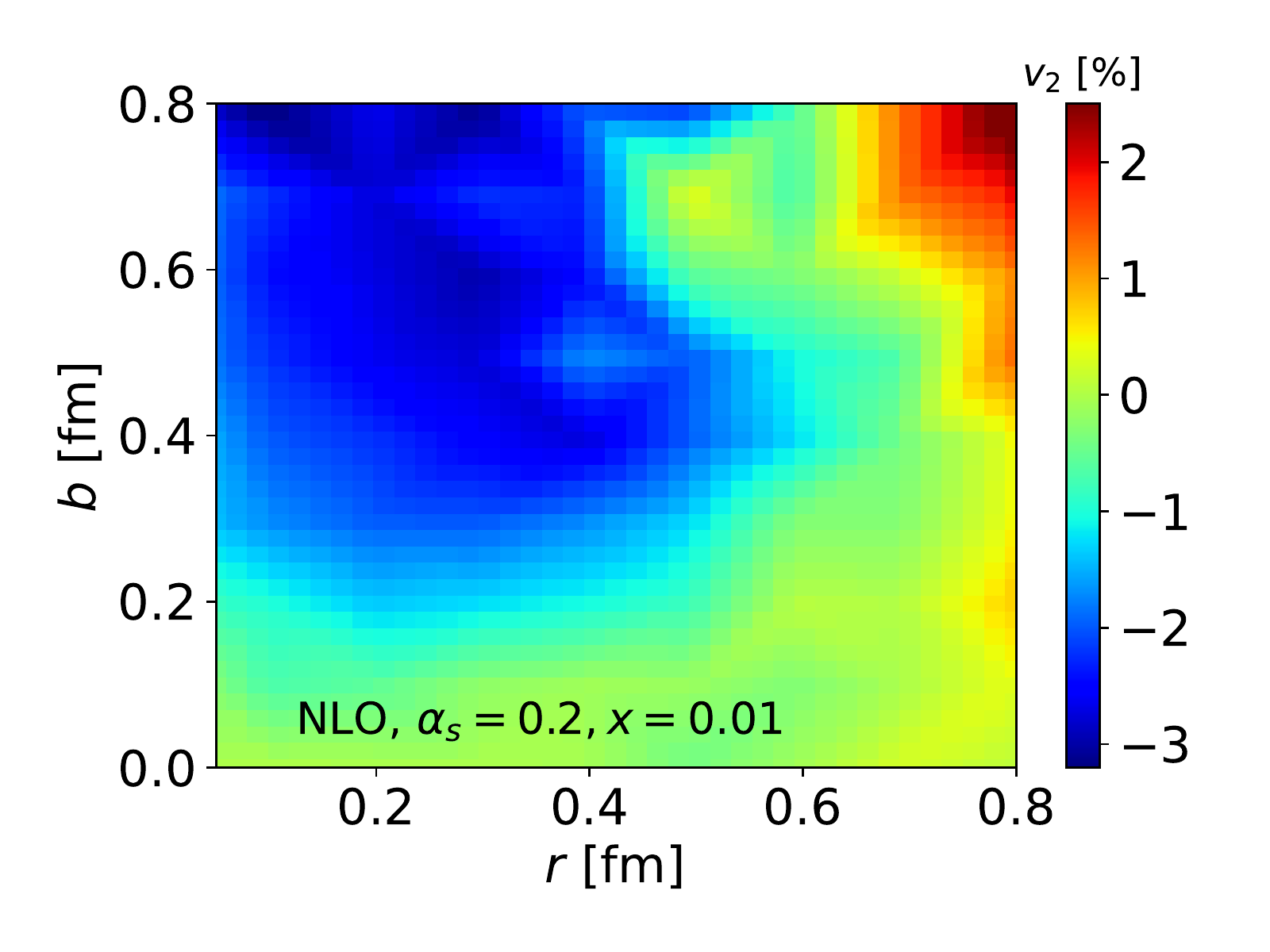}
\caption{$v_2=\langle\cos 2\theta\rangle$ from the dipole scattering
  amplitude at NLO in the impact parameter vs.\ dipole size plane at $x=0.01$. The smallest dipole size at the left edge of the plot is $r_\text{min}=0.05~\fm.$ }
\label{fig:dipole_v2_heatmap_001}
\end{figure}

The amplitudes $v_2= \langle\cos2\theta\rangle$ of the azimuthal
modulation in the $b$ vs.\ $r$ plane at $x=0.1$ and at $x=0.01$ are  shown in
Figs.~\ref{fig:dipole_v2_heatmap_01} and~\ref{fig:dipole_v2_heatmap_001}. 
Here, we observe a rather mild modification of the azimuthal asymmetry with decreasing $x$.
Although the NLO effects dominate the scattering amplitude at $x=0.01$ (as shown above), the 
azimuthal amplitude $v_2$ does not depend strongly on the momentum fraction $x$, and consequently the NLO corrections have a small effect on $v_2$.
The fact that $v_2$ is approximately independent of $r$ at $r\lesssim 0.5$~fm can be clearly seen from these figures. We find $v_2<0$ in the region of $r, b$ where the perturbative calculation is reliable. In particular, the dependence on the dipole size $r$ is completely different to what is obtained from  the impact parameter dependent MV model shown in Eq.~\eqref{eq:N_MV} and numerically studied in Ref.~\cite{Mantysaari:2020lhf}.

%-------------------------------------------------------------

\section{Summary and Outlook}

In this Letter, we have computed and shown the behavior of color charge
correlations (at quadratic order in $\rho^a$) in a proton on the light
cone. At LO the proton is approximated by a three quark state with
a wave function that is consistent with its empirically observed structure
at large $x$~\cite{Schlumpf:1992vq,Brodsky:1994fz}, which we refine by
including NLO corrections due to emission or exchange of a perturbative
gluon. Our results apply in a regime of moderately high energy,
perhaps corresponding to light cone momentum fractions of
$x=0.01 \ldots 0.1$, where scattering is approximately eikonal but
still far from the unitarity limit.

We have emphasized that the correlator exhibits non-trivial behavior
at large momentum transfer or central impact parameters. In that
regime, it does not merely trace the behavior of a transverse proton
shape function but displays repulsive correlations due to the ``cat's
ears diagram'' at leading order, which persist at next-to-leading
order. Furthermore, the color charge correlation function depends not
only on the impact parameter $\vec b$ but also on the relative
transverse momentum $\vec q_{12}$ of the two gluon probes, and on
the angle made by $\vec b$ and $\vec q_{12}$. This is in contrast to
the McLerran-Venugopalan (MV) model~\cite{McLerran:1993ni,McLerran:1993ka},
which is used extensively in the literature.

Furthermore, from the two-point color charge correlator, we have
computed the dipole scattering amplitude $N(\vec r, \vec b; x)$ in the
two gluon exchange approximation. This includes the contributions from
the soft and collinear singularities without restriction to the
leading logarithmic approximation. We observe a strong amplification
of color charge density fluctuations with decreasing $x$; from $x=0.1$
to $x=0.01$ the dipole scattering amplitude $N(\vec r, \vec b; x)$
increases by factors of $\sim 3$ (for $\as=0.2$). We also find that the
azimuthal anisotropy of the dipole scattering amplitude is affected
significantly by the angular dependence of the color charge
correlations. We observe a behavior of $\langle\cos 2\theta\rangle$
which, in some ranges of impact parameter and dipole size, differs
substantially from expectations based on isotropic color charge
correlators proportional to the proton profile function $T_p(b)$ \cite{Altinoluk:2015dpi,Iancu:2017fzn,Mantysaari:2019csc,Mantysaari:2020lhf}. 

Our computations also provide initial conditions for
Balitsky-Kovchegov (BK) evolution of the dipole scattering amplitude to
lower $x$~\cite{Balitsky:1995ub,Kovchegov:1999yj,Ducloue:2019ezk}; in particular, for
impact parameter dependent evolution~\cite{GolecBiernat:2003ym,Berger:2010sh,Berger:2011ew,Berger:2012wx,Cepila:2018faq, Bendova:2019psy,Cepila:2020xol,Bendova:2020hkp}. This initial
condition depends not only on the impact parameter and the dipole
vectors but also on their relative angle, and on the light-cone
momentum fraction $x$ in the {\em target}.  
In the future, proton ``imaging'' 
in
the regime of small and moderate $x$
performed at the EIC~\cite{Boer:2011fh,Accardi:2012qut,Aschenauer:2017jsk,AbdulKhalek:2021gbh},
 at other future nuclear DIS facilities~\cite{Agostini:2020fmq,Anderle:2021wcy} or in ultra peripheral collisions~\cite{Bertulani:2005ru,Klein:2019qfb} at the LHC, will further constrain the proton
light cone wave function, and the dipole scattering amplitude which we
relate to it.

We close with a brief outlook. We have already mentioned the
successful phenomenology that emerged from the picture of a
fluctuating proton substructure~\cite{Mantysaari:2016ykx,Mantysaari:2016jaz,Mantysaari:2017dwh,Mantysaari:2017cni, Albacete:2016pmp, Albacete:2016gxu,Cepila:2016uku,Cepila:2017nef,
  Albacete:2017ajt,Traini:2018hxd,Mantysaari:2018zdd,Moreland:2018gsh,Mantysaari:2019jhh} (see \cite{Mantysaari:2020axf} for a review). It will be very interesting to reformulate these
approaches so that the ensemble of quark and gluon configurations in
the proton would be determined by its light cone wave function at NLO
rather than be based on geometric pictures.

\section*{Acknowledgements}
We thank A.~Kovner, T. Lappi, F.~Salazar and V.~Skokov for discussions.

This work was supported by the Academy of Finland, projects 314764
(H.M) and 1322507 (R.P). H.M.\ is supported under the European Union’s
Horizon 2020 research and innovation programme  STRONG-2020 project (grant agreement no.
824093), and R.P.\  by the European Research Council grant agreement no. 725369. A.D.\ thanks the 
US Department of Energy, Office of Nuclear Physics, for support
via Grant DE-SC0002307.
The content of this article does not reflect the official
opinion of the European Union and responsibility for the information
and views expressed therein lies entirely with the authors. Computing
resources from CSC – IT Center for Science in Espoo, Finland and from
the Finnish Grid and Cloud Infrastructure (persistent identifier
\texttt{urn:nbn:fi:research-infras-2016072533}) were used in this
work.

\bibliographystyle{JHEP-2modlong.bst}
\bibliography{refs}

\end{document}